\documentclass {article} %, svglov3.clo   article  

\usepackage {amsfonts,citesort} % in bibliography for unsrt

\begin{document}

\author{J. Brian Pitts \\ 
100 Malloy Hall \\ % Department of Philosophy \\ 
University of Notre Dame \\ Notre Dame,  Indiana 46556, USA \\ jpitts@nd.edu}

\sloppy

\title{Gauge-Invariant Localization of  Infinitely Many Gravitational Energies from All Possible Auxiliary Structures} 

\maketitle

\abstract{ The problem of finding a  covariant expression for the distribution and conservation of gravitational energy-momentum dates to the 1910s.   A suitably covariant infinite-component localization is displayed, reflecting Bergmann's realization that there are infinitely many conserved gravitational energy-momenta.  Initially use is made of a flat background metric (or rather, all of them) or connection, because  the desired gauge invariance properties are obvious.  Partial gauge-fixing then yields an appropriate covariant quantity without any background metric or connection; one version is the  collection of pseudotensors of a given type, such as the Einstein pseudotensor, in \emph{every} coordinate system.  This solution to the gauge covariance problem is easily adapted to any pseudotensorial expression (Landau-Lifshitz, Goldberg, Papapetrou  or the like) or to any tensorial expression built with a background metric or connection.  Thus the specific functional form can be chosen on technical grounds such as relating to Noether's theorem and yielding expected values of conserved quantities in certain contexts and then rendered covariant using the procedure described here. The application to angular momentum localization is straightforward. Traditional objections to pseudotensors are based largely on the false assumption that there is only one gravitational energy rather than infinitely many. }  %c. 193 words, okay for GRG 

Short title ``Gauge-Invariant Localization of  Infinitely Many Gravitational Energies''

PACS numbers 04.20.Cv 	Fundamental problems and general formalism,
04.20.Fy 	Canonical formalism, Lagrangians, and variational principles,
11.30.-j 	Symmetry and conservation laws


%%%%%%%%%%%%%%%%%%%%

\section{Introduction}

The problem of finding a  covariant expression for the distribution and conservation of gravitational energy-momentum for General Relativity dates to the 1910s.
Einstein took the requirement that the gravitational field equations alone entail energy-momentum conservation as a criterion for finding his field equations in his process of discovery \cite{EinsteinEntwurf,NortonField,Janssen,JanssenRenn}; ironically, it was widely concluded that the final  theory lacked any local conservation law for energy-momentum. The equation 
\begin{equation} 
\nabla_{\mu} T^{\mu\nu} =0,
\end{equation}
though a consequence of Einstein's equations, is a balance equation, not a conservation equation, because the covariant divergence of a rank 2 tensor (with any index placement and density weight) cannot be written using a coordinate divergence, as is required for integral conservation laws.  
 Gravitational energy-momentum has been reviewed on several occasions  \cite{FletcherConservation,TrautmanConserve,GoldbergReview,SzabadosReview}. While there is no difficulty in writing down quantities satisfying local conservation laws (in the sense of a coordinate divergence), there seem to be too many  expressions without the anticipated interconnections \cite{GravesEnergy}.  More specifically, it has been expected that there ought to be a (10- or 16-component) tensor, geometric object, or other suitably covariant expression that describes the local distribution of gravitational energy-momentum, and yet evidently  there is not one.  Pseudotensorial answers go back to the Einstein's work in 1916 \cite{EinsteinFoundation}, while objections to them from Schr\"{o}dinger and from Bauer appeared in 1918 \cite{SchrodingerEnergy,BauerEnergy,Pauli,Cattani}.  Later developments included  the introduction of additional background structures, such as a flat background metric \cite{Rosen1,RosenAnn,Cornish1},
 an orthonormal tetrad \cite{MollerAnnals,MollerRadiation}, or a flat connection \cite{SorkinFlux,FatibeneFrancaviglia}.  While the introduction of such further structures has achieved  tensorial form with respect to coordinate transformations, this result has always come at the cost of introducing a new sort of  gauge dependence, because the choice of specific background metric, tetrad, or connection lacks physical meaning and yet affects the results.  The introduction of additional structures appears simply to move the lump in the carpet, not to flatten it out. Though new background structures %(such as spinors)
 continue to be introduced, the inductive lesson only gets stronger that the gauge dependence problem is not resolvable in such a fashion \cite{SzabadosReview}.  In this respect it is unclear that much has been gained beyond the original dependence of pseudotensors on coordinates found in the 1910s.

The solution to the problem of gauge dependence, briefly, is to take \emph{all possible auxiliary structures of a given type together}.  Thus, for example, the collection of all flat background metrics does not depend on the choice of any particular background metric.   Changing the flat background metric from one specific example to another merely leads to another member of the same collection. Looking for some finite-component expression that is covariant under a change of the background metric, though traditional, is a mistake.   Similar remarks hold for tetrads, connections, and even coordinate systems.  Indeed the cases of background metrics, background connections, and coordinate systems seem closely related, while the tetrad case differs and so will not be discussed much here. Its introduction of a gratuitous local Lorentz group is a major disadvantage, and it is in fact not required for spinors, as will appear below.  

Some authors, especially those who emphasize how different General Relativity is from other field theories rather than how similar it is, have tried to make the best out of the apparent non-existence of gauge-invariant gravitational energy localization.  Thus the question has been rejected as inappropriate, as shown by the equivalence principle \cite{MTW}: ``[a]nybody who looks for a magic formula for `local gravitational energy-momentum' is looking for the right answer to the wrong question.'' \cite[p. 467]{MTW}  However, this is an \emph{ad hoc} move.  Noether's theorems do not care about the equivalence principle; they simply give results in any coordinate system \cite{BradingConserve}.  Rather than criticizing the results  of Noether's theorem in terms of preconceived notions of invariance and then mysteriously invoking a principle irrelevant to Noether's theorem  to reduce the puzzlement over the lack of an invariant energy complex, it is preferable to learn from the results of Noether's theorem that there is a broader notion of invariance suited to the existence of infinitely many distinct conserved energies.   There is no reason to expect the components of a pseudotensor to transform into each other once the vast multitude of gravitational energy-momenta is recognized.  The importance of considering messy  mathematical details rather than relying on geometrical shortcuts and picture-thinking is increasingly being recognized both in  technical General Relativity literature \cite{Kiriushcheva,KiriuschchevaMyths} and the foundations of physics \cite{BrownPhysicalRelativity}.

%%%%%%%%%%%%%%%%%%%%%%%%%
\section{Infinite-Component Geometric Objects}

At this stage it will be helpful to introduce the notion of an infinite-component geometric object.  An old standard definition of a  geometric object  (slightly streamlined for physicists' use in local field theories) by Trautman assumes a finite number $N$ of ordered components:
\begin{quote} Let $X$ be an $n$-dimensional differentiable manifold.\ldots 

Let $p \in X$ be an arbitrary point of $X$ and let $\{x^a\}, \{x^{a^\prime} \}$ be two systems of local coordinates around $p.$ A \underline{geometric object field} $y$ is a correspondence $$ y: (p, \{x^a\}) \rightarrow (y_{1}, y_{2},\cdot \cdot \cdot \; y_{N}) \in R^N$$ which associates with every point $p \in X$ and every system of local coordinates $\{x^a\}$ around $p$, a set of $N$ real numbers, together with a rule which determines $(y_{1^\prime}, \cdot \cdot \cdot \; y_{N^\prime})$, given by $$ y: (p, \{x^{a^\prime} \}) \rightarrow (y_{1^\prime}, \cdot \cdot \cdot \; y_{N^\prime}) \in R^N$$
 in terms of the  $(y_{1}, y_{2},\cdot \cdot \cdot \; y_{N})$ and the values of [\emph{sic}] $p$ of the functions and their partial derivatives which relate the coordinate systems $\{x^a\}$ and $ \{x^{a^\prime} \}.$\ldots  The $N$ numbers  $(y_{1}, \cdot \cdot \cdot y_{N})$ are called the \underline{components} of $y$ at $p$ with respect to the coordinates $\{x^a\}$. \cite[pp. 84, 85]{Trautman} \end{quote}  
(In more modern-style literature, geometric objects have turned into natural bundles \cite{NijenhuisYano,FatibeneFrancaviglia}.)
The infinite-component entity needed for present purposes has the same cardinality as the set of all flat metric tensors, the set of all vector fields, or the set of all coordinate systems (with some continuity assumptions), so imposing an order is an unattractive prospect. No ordering is required to gather the components into a set, however. %; one still wishes to be able to identity particular components with some kind of labeling scheme, however.  
One may therefore take an infinite-component geometric object to be analogous to a geometric object of the familiar sort, but with an infinity of components collected into a set. % and labeled. 
 An example of an infinite-component geometric object is the set of all flat metric tensors.  Using the universal quantifier $\forall$ (``for all''), one can write this object as  \begin{equation}
\{ (\forall \eta_{\rho\sigma})  \;  \eta_{\rho\sigma}  \}.
\end{equation} (Here one can read the Greek indices as abstract indices.)
In this case each element is a coordinate tensor and hence a geometric object in the usual sense, but that feature is not guaranteed in general.

%%%%%%%%%%%%%%%%%%%%%

\section{Infinite-Component Covariant Density in Terms of All Flat Backgrounds}

It might seem that achieving covariance of the gravitational energy-momentum distribution by letting it depend on \emph{all possible} flat background metrics (or other auxiliary structures) would give a baroque construction without physical meaning.  That, however, is  wrong, primarily because it reflects the almost universal but usually tacit assumption that there  ought to be just one gravitational energy-momentum (with 10 or perhaps 16 components).  This assumption of uniqueness is especially clear in treatments by Goldberg \cite{GoldbergReview}, Faddeev \cite{FaddeevEnergy} and  Szabados \cite[section 3.1.3]{SzabadosReview}. Faddeev writes, ``The energy of the gravitational field is not localized, i.e., a uniquely defined energy density does not exist.'' \cite{FaddeevEnergy} While stated with special clarity in some cases, the assumption of uniqueness is implicit almost everywhere in the literature in the expectation that a pseudotensorial expression (perhaps Einstein's) in one coordinate system ought ideally to be related by a transformation law to that pseudotensor in another coordinate system in order to have the intended physical meaning of representing gravitational energy-momentum density.  
  This expectation of uniqueness makes sense if, as in other theories, there is only one energy in General Relativity.  It has been known at least since 1958 due to Bergmann and Komar, however, that there are \emph{infinitely many} gravitational energies, and that any vector field generates one  \cite{BergmannConservation,KomarConservation}.  Some of them might be zero; for example, a vector field derived by index-raising from an exact covector has vanishing Komar energy density. (The resulting Komar energies are unsatisfactory \cite{PetrovKatz2}, so there is reason to expect the energies to depend on more than just a single vector field and the metric.) Some of the energies might plausibly regarded as faces of a single energy, such as if a Lorentz or affine transformation relates them.  But the point remains that there are a great many \emph{different} gravitational energy-momenta, uncountably infinitely many, far more than one naively expected, and \emph{any} vector field (subject to some restrictions on differentiability, \emph{etc.}) yields one. Why can't they all be real? (The assumption of uniqueness has been so widespread, however, that even Komar went on to look for restrictions on the vector field that his formalism required with the goal of achieving or approaching uniqueness \cite{KomarConservation,KomarAsymptotic,KomarPositive}.) Thus there is no reason whatsoever to expect distinct conserved quantities to behave mathematically as though they were just faces of one (finite-component) conserved quantity; the paradox dissolves. If a transformation law relating the components of the Einstein pseudotensor existed, then its components in one coordinate system would determine its components in all coordinate systems, thus implying that there was only one energy, a known falsehood.  
 An arbitrary vector field also generates a coordinate transformation, whether of the familiar infinitesimal form (\emph{e.g.}, \cite{Anderson}) or the finite form \cite{Nijenhuis,Grishchuk,PetrovChapter}.  An arbitrary coordinate transformation will convert one flat metric into any other, so using all flat background metrics (or all flat connections or all coordinate systems) plausibly gives the right number of gravitational energy-momentum densities.

Let $t^{\mu}_{\nu}$ be one's favorite gravitational energy-momentum tensor, or related to it by index lowering and perhaps density-reweighting with the flat  metric $\eta_{\mu\nu}.$ This expression presumably is chosen based  on technical considerations involving getting the expected values for integrated conserved quantities in suitable contexts, relation to Noether's theorem, and the like.  A good candidate  is due to Joseph Katz,  Ji\v{r}\'{i} Bi\v{c}\'{a}k and Donald Lynden-Bell \cite{KatzBicakLB,KatzEnergy,PetrovChapter}. Or perhaps the appropriate form depends on the boundary conditions \cite{NesterQuasiPseudo,NesterQuasi}.\footnote{As a referee notes, the work of Nester and collaborators and the proposal here have in common a tendency to find meaning in infinite ambiguity rather than to reject it.} 
Whatever the specific form, this gravitational energy-momentum tensor $t^{\mu}_{\nu}[g_{\alpha\beta}, \eta_{\rho\sigma}]$ for a given curved metric $g_{\alpha\beta}$ is some functional of the curved metric $g_{\alpha\beta}$ and a flat metric tensor $\eta_{\rho\sigma},$ depending on their values and maybe one or two partial derivatives. (Alternatively, a mere flat connection can be used.) 
Instead of taking partial derivatives, one can take covariant derivatives using the flat connection built out of  $\eta_{\mu\nu},$ yielding a manifestly tensorial but still gauge-dependent expression.   When General Relativity is formulated with a background metric, the action has two invariances, one under changes of coordinates and one under gauge transformations.  The latter transformations alter the mathematical relationship between $g_{\mu\nu}$ and $\eta_{\mu\nu}.$ For this reason $t^{\mu}_{\nu}$ is tensorial with respect to coordinate transformations, but gauge-variant under gauge transformations  \cite{Grishchuk,Zel1,Grishchuk90,BrazilLocalize,PetrovChapter}.  
For finite one-parameter transformations, one can write coordinate transformations as 
\begin{eqnarray}
	g_{\sigma\rho} \rightarrow e^{\pounds_{\xi}} g_{\sigma\rho},
u \rightarrow e^{\pounds_{\xi}}   u,
	\eta_{\mu\nu} \rightarrow e^{\pounds_{\xi}} \eta_{\mu\nu}, \label{coord}
\end{eqnarray} where $u$ stands for any bosonic matter fields.  (Spinors will be discussed in the next section.)   This transformation induces the same Lie-Taylor series for the connection, using the commutativity of Lie and partial derivatives and the Leibniz rule for the Lie derivative \cite{Nijenhuis,Yano,NullCones1}.By contrast gauge transformations are written as 
\begin{eqnarray}
	g_{\sigma\rho} \rightarrow e^{\pounds_{\xi}} g_{\sigma\rho},
u \rightarrow e^{\pounds_{\xi}}   u,
	\eta_{\mu\nu} \rightarrow \eta_{\mu\nu}, \label{gauge}
\end{eqnarray} which leave the flat metric (and connection) alone.  If one wishes, one can combine a gauge transformation with a coordinate transformation in the `opposite direction' to yield a modified gauge transformation that alters the background metric $\eta_{\mu\nu}$ while  leaving $g_{\mu\nu}$ and any matter fields $u$ alone \cite{SliBimGRG}. 
Clearly the set of all flat metrics 
$ \{ (\forall \eta_{\rho\sigma})  \;  \eta_{\rho\sigma}  \}$
is gauge-invariant: a gauge transformation changes one flat metric into another, but the set as a whole is unchanged.

One can now write down the infinite-component covariant expression for the distribution of gravitational energy-momentum.  It is obtained by simply collecting all the energy-momentum tensors together for the various background metrics:
\begin{equation}
\{ (\forall \eta_{\rho\sigma})  \;  t^{\mu}_{\nu}[g_{\alpha\beta}, \eta_{\rho\sigma}] \}.
\end{equation}
This expression does not depend on the choice of any particular flat background metric, so it is gauge-invariant.  Each element is a coordinate tensor;  the whole collection is gauge-invariant
although no part of it is.   Feeding the gauge-invariant set $\{ (\forall \eta_{\rho\sigma})  \;  \eta_{\rho\sigma}  \}$ of all flat
background metrics  into the stress energy tensor formula gives a gauge-invariant set
of energy-momentum tensors. A gauge transformation turns a specific element $ t^{\mu}_{\nu}[g, \eta_1] $ into another element $ t^{\mu}_{\nu}[g, \eta_2],$ but the set is unchanged.  Each element of the set is covariantly conserved with respect to the torsion-free connection induced by its own flat metric, due to Einstein's field equations: 
\begin{eqnarray}  \partial_{1\mu} t^{\mu}_{\nu}[g, \eta_1] = 0, \partial_{1\mu} \eta_{1\alpha\beta} \equiv 0, \nonumber \\
\partial_{2\mu} t^{\mu}_{\nu}[g, \eta_2] = 0, \partial_{2\mu} \eta_{2\alpha\beta} \equiv 0, \nonumber 
\end{eqnarray}
 \emph{etc.} The generalization to the use of a mere flat background connection is immediate.  Note that if (\emph{per impossible}) there were a nonzero tensorial and gauge-invariant expression with only 10 (or 16) components, as many have wished, then it could represent only a single energy, rather than the infinitely many that Bergmann and Komar taught us to expect.  When Bergmann and Anderson said that gravitational energy-momentum did not form a geometric object \cite{BergmannConservation,Anderson}, it was assumed that the geometric object would have finitely many components.  Instead there is an infinite-component geometric object. 
The  expression  given here has infinitely many components in two senses:  a somewhat trivial sense due to its availability in any coordinate system  and the nontrivial sense due to the use of every flat background metric (so it describes infinitely many energies). A similar gauge-invariant collection could be obtained, for example, by raising an index and reweighting with the relevant flat metrics to get 
\begin{equation}
\{ (\forall  \eta_{\rho\sigma})  \;  \sqrt{-\eta} t^{\mu\nu}[g_{\alpha\beta}, \eta_{\rho\sigma}] \}.
\end{equation}    
Obtaining a gauge-invariant quantity by collecting together the result for every gauge bears some  resemblance to the technique of group averaging \cite{MarolfGroup}, but in this case one merely collects the pieces together into a set rather than adding them up.

Constructing  a gauge-invariant set by collecting together an expression in every gauge works even if one quantifies only over all elements satisfying some suitable condition, perhaps some inequalities restricting the allowed coordinates \cite{Hilbert2,MollerBook} or allowed bimetric gauges \cite{NullCones1}:  the gauge-invariant collection is found by collecting the complexes for all the allowed gauges or coordinates.  One then has only a (Brandt) groupoid, not a group, of gauge transformations \cite{NullCones1}: the allowed transformations depend on the configuration, so not every pair of elements can be multiplied.

 The problem of treating integral conservation laws is not addressed here. However, it seems evident that having both coordinate freedom and gauge freedom available \emph{via}  the use of a background metric or connection would be helpful in permitting the coordinates to be adapted to the integration hypersurface while retaining gauge freedom.   % Katz 

Given that the Hilbert action gives the wrong (Komar) conserved quantities \cite{PetrovKatz2}, an alternative dependent on a coordinate system or background metric or connection is required \cite{FaddeevEnergy,HawkingHorowitz,FatibeneFrancaviglia}.  While any such action is gauge-dependent, one can obtain a gauge-invariant multi-action principle by feeding all possible background structures into the Lagrangian density, thereby obtaining an infinite-component Lagrangian density.  The equivalence of the field equations from the many Lagrangians should render this procedure innocuous at least at the classical level.

%%%%%%%%%%%%%%%%%%%%%
\section{Spinors as Almost Geometric Objects}

Given the most common ways of treating spinor fields, it is not obvious how gravitational energy localization in the form proposed here would work.  
Introducing an orthonormal basis and treating spinors as coordinate scalars is a standard move (when an orthonormal basis exists \cite{ChoquetDeWitt2}).  
M{\o}ller's orthonormal tetrad formalism was motivated in part by its supposed necessity to accommodate spinor fields \cite{MollerAnnals}.   The  local Lorentz group  introduced in the tetrad formalism  \cite{MollerRadiation} seems quite unhelpful for localizing gravitational energy, however, even if one accepts all the tetrads at once.  Whereas the background metrics or background connections are closely related to the coordinate transformation freedom that is already present and ineliminable from the manifold, the local $O(3,1)$ group apparently bears no such relation. Thus the gauge invariant energy localization scheme presented here seems potentially inapplicable or at best purely formal in the presence of spinors.  

  Fortunately it is not the case that a tetrad is necessary for spinors, contrary to widely held opinion.  Thus the local Lorentz group is gratuitous not only in relation to gravitational energy localization, but also in relation to coupling spinors to a curved metric.  The tetrad formalism and local Lorentz group follow only if one insists on a linear coordinate transformation law for spinors as opposed to a nonlinear one \cite[p. 234]{GatesGrisaruRocekSiegel} \cite{OPspinor}.  It is possible to include spinor fields almost like tensors in the Ogievetsky-Polubarinov-Bilyalov formalism \cite{OPspinor,OP,BilyalovSpinors}. 
The spinor and  the metric  together form  a nonlinear geometric object $\langle g_{\mu\nu}, \psi  \rangle$  \cite{OPspinor,OP,BilyalovSpinors} (up to a sign for the spinor part), % \cite{Siwek}),
 with mild restrictions on the admissible coordinates  to distinguish the time coordinate from the spatial coordinates.  (The inequalities restricting the coordinates serve the same purpose as Bilyalov's matrix $T$ that interchanges two coordinates \cite{BilyalovConservation} to get time listed first.) 
The nonlinearity is due to the fact that the new components of the spinor depend not only (linearly) on the old spinor components, but also on the metric in a nonlinear fashion \cite{OPspinor}.  
Nonlinear geometric objects in classical differential geometry, which were studied briefly in the 1950s-60s \cite{SzybiakLie,SzybiakCovariant,Yano}, turn out to be basically a
special case of the nonlinear group representations that particle physicists started studying in the  1960s \cite{IshamSalamStrathdee,ChoFreund}.  $\langle g_{\mu\nu}, \psi  \rangle$ is a nonlinear representation (up to a sign for the spinor) of the general coordinate transformation group, or a sufficiently large subgroupoid thereof, which is linear for the Poincar\'{e} subgroup, and indeed for the 15-parameter conformal group, the stability group. Roughly  and locally speaking, the Ogievetsky-Polubarinov-Bilyalov formalism resembles the tetrad-spinor formalism with the tetrad in the symmetric gauge.  However,  the symmetric square root of the metric makes sense on any manifold with a  metric (with mild coordinate restrictions), unlike an orthonormal basis. 
 Thus spinors, as treated in (\cite{OPspinor,OP,BilyalovSpinors}), require some  technical modifications, but do not require treatment fundamentally different from tensors and more general geometric objects in classical differential geometry. In particular, Lie and covariant differentiation are well defined for $\langle g_{\mu\nu}, \psi  \rangle$  \cite{OP,BilyalovSpinors}, though not for the spinor separately, just as one expects  for nonlinear geometric objects \cite{SzybiakLie,SzybiakCovariant,Yano}; such invariant derivatives need only the coordinate transformation behavior near the identity. 
This excursus on spinors shows that the gravitational energy localization proposed here also applies to spinors, which is not at all obvious for some well known  spinor formalisms.

%%%%%%%%%%%%%%%%%%%%%%%%%%%%%%%%%%%%%%%

\section{Infinite-Component Covariant Energy Density in Terms of All Coordinates and One Metric}

The use of a background metric or connection, or rather, of the whole collection thereof, is actually not essential to the technique of getting a gauge-invariant infinite component localization of gravitational energies. The use of a background metric or connection has the virtue that it manifestly has every sort of invariance that one would expect---both tensoriality under coordinate transformations and covariance under gauge transformations.  It is initially somewhat less clear what one should expect in a formalism with no background metric.  Fortunately one can gauge-fix the formalism above with a flat background metric or connection to find out.  I will ignore global issues by  pretending that all coordinate charts are defined everywhere.  One ought to globalize the results using bundles, but the basic idea will be clear without such techniques.  Globalizing the results for topologically nontrivial space-times might be nontrivial \cite{Misnerg00,ThirringForms}, but recalcitrant difficulties might be features of gravitational energy rather than limitations of the formalism at hand. (The use of a background metric or connection provides a more globally robust formalism \cite{FatibeneFrancaviglia}.) 

% likely energy only relevant in space-times that reduce to Minkowski as parameter varies:  \cite{DeserBrillVariational,DeserBrillPositive}

One convenient  gauge fixing takes the bimetric formalism above and dispenses with the flat background metric tensors  by choosing (for example) Cartesian coordinates for each flat metric separately.  Thus each flat metric tensor   $\eta_{\mu\nu}$ in the set $\{ (\forall \eta_{\rho\sigma})  \;  \eta_{\rho\sigma}  \}$  is downgraded to a \emph{matrix} \begin{equation} \eta_{MN}=diag(-1,1,1,1) \end{equation} and its resulting connection is downgraded to a three-index entity with only vanishing components, which can be ignored.  Now the former coordinate freedom   (\ref{coord})
 is destroyed, but the former gauge freedom (\ref{gauge})
 is formally converted into coordinate freedom (which has no effect on the numerical matrix $\eta_{MN}$).  The new coordinate freedom is still gauge freedom in the sense of Dirac-Bergmann constrained dynamics \cite{Sundermeyer}).  In a chart one has one's favorite pseudotensor 
$ t^{\mu}_{\nu}[g_{\mu\nu}, \eta_{MN}],$ where the expression $g_{\mu\nu}$  now means the coordinate components of the curved metric. Using Einstein's field equations, the pseudotensor $ t^{\mu}_{\nu}[g_{\mu\nu}, \eta_{MN}]$ (or $ t^{\mu}_{\nu}[g_{\mu\nu}]$) is conserved in the sense of having vanishing \emph{coordinate} divergence 
\begin{equation} \frac{\partial}{ \partial x^{\mu} } t^{\mu}_{\nu}[g_{\mu\nu}, \eta_{MN}] =0 \end{equation}
in every coordinate system. A vanishing coordinate divergence is just what one needs to obtain an integral conservation law \cite{Anderson}.  The gauge-invariant infinite-component gravitational energy-momentum distribution is just a certain pseudotensor in every coordinate system $U$:
\begin{equation}
\{ \forall U \; t^{\mu}_{\nu}[g_{\mu\nu}, \eta_{MN}]  \}.
\end{equation}
The curved metric thus appears in all possible coordinate systems. 
This expression for the localization of gravitational energies has infinitely many components in a nontrivial sense:  each coordinate system picks out a distinct conserved energy.  The distinctness depends on the fact that the expression $t^{\mu}_{\nu}$ is not a tensor (or other geometric object \cite{BergmannConservation,Anderson}). The components of a tensor or  any geometric object with respect to all coordinate systems give infinitely many faces of the same entity, but here we have infinitely many distinct entities, each appearing in its own adapted coordinate system.

If one previously used a flat background connection only, rather than a flat background metric, then the auxiliary matrix    $\eta_{MN}$ is not present.  Some pseudotensors depend on the matrix $\eta_{MN},$ such as Papapetrou's \cite{Papapetrou,LeclercNoether}, while others do not, such as Einstein's. Some time ago Goldberg found a family of energy-momentum pseudotensors and a family of angular momentum complexes, but the preferred versions lacked the simple relationship for which one might have hoped \cite{GoldbergConservation}.    If one admits a background metric as a reference configuration, then many more options are available and this problem disappears \cite{ChangNesterChen}.

One might think that, in a mature subject such as differential geometry, every mathematical entity worth using would have a name and that its name would reflect its usefulness.   As it turns out, there is a useful mathematical entity that does not have a name based on the reason that it is useful, and pseudotensors are an example of it.  In older literature on geometric objects, one encounters at a preliminary stage the concept of ``object,'' used as a technical term (for example, \cite{Nijenhuis,KucharzewskiKuczma}). Thus Nijenhuis writes \cite[p. 28]{Nijenhuis}:  \begin{quote}
[t]he definition of the geometric object goes via the object:  an \underline{object}  at a point $P$ of $X_n$ is a correspondence between all coordinate systems defined for $P$ and sets of $N$ numbers, such that with each coordinate system there is associated one such set of numbers, called the \underline{components} of the object with respect to the coordinate system.  An object \underline{field}  in a region $R$ of $X_n$ is a correspondence between all coordinate systems defined in subregions of $R$, and sets of $N$ functions, such that with each coordinate system there is associated one such set of functions, defined and analytic in the region in which the coordinate system is defined.  The values taken by the functions at a point are the components of the object at that point with respect to the coordinate system to which the functions belong.\end{quote}
An object (field) counts as a geometric object (field) if and only if a there is a  transformation law relating the components in the various (overlapping) coordinate systems.  It is evident that the collection of components of one's favorite pseudotensor in every coordinate system forms an object in Nijenhuis's sense, but not a geometric object.

The usual attachment to geometric objects (including tensors) is due in part  to the unity imposed by the transformation law.  Without a transformation law, the components of an object in different coordinate systems might have nothing to do with each other, apart from the stipulation that they are components of the object in question.   The different sets of components pick out distinct entities, rather than representing the same entity relative to different conventional choices of coordinates.  Not all geometric objects are physically interesting, however.  Some of them represent things in physical theories, while others are merely mathematical collections of numbers bearing an interesting formal relation of equivalence.  It is clear, then, that having a  transformation law in itself is not what makes a geometric object of interest.  What the transformation law does is ensure that the seemingly disparate components of an object are in fact equivalent, so if one set of components has physical meaning, then the other sets of components have that same physical meaning.  The components of an object can have a physical meaning or not, and they can be interrelated by a transformation law (yielding a geometric object) or not.  It is important to realize that these questions are independent, so that one could potentially have a physically meaningful geometric object, a physically meaningless geometric object, a physically meaningless non-geometric object, and, most importantly for present purposes,  even a physically meaningful non-geometric object.  Pseudotensors are an example of the latter.  If the idea of a physically meaningful non-geometric object is difficult, a formal move below will show it to be numerically equal to an infinite-component geometric object built by taking all the natural coordinate bases as auxiliary objects.

It will be useful to compare and contrast the set of pseudotensor components with respect to every coordinate system with the set of Einstein tensor components with respect to every coordinate system and the set of metric tensor components with respect to every coordinate system.  For a geometric object,  one has a set of components at every point (where defined) in every chart, and also a transformation rule to infer one set of components from another, as appeared above  \cite{Nijenhuis,Trautman}.
Thus the components of a geometric object form a natural kind mathematically:  they constitute faces of one and the same entity by virtue of being interrelated by a coordinate transformation law.  

For the Einstein tensor $G^{\mu\nu}$ or the metric tensor  $g_{\mu\nu},$ one has a further sort of unity in terms of physical meaning.  For the Einstein tensor, the physical meaning is displayed in  a recipe for constructing the components of the Einstein tensor in a coordinate system from the components of the metric and its partial derivatives in that same coordinate system.  Because of the tensor transformation law, the various sets of components of $g_{\mu\nu}$  and of $G^{\mu\nu}$ are physically and mathematically equivalent; they are just faces of the same entity, the metric tensor or the Einstein tensor, respectively. 
The components of the Einstein tensor form a natural kind in two senses,  mathematically by virtue of the tensor transformation law and physically by virtue of being constructed from the metric by the same recipe in every coordinate system. Likewise the components of the metric tensor form a natural kind not only  mathematically by the tensor transformation law, but also physically by virtue of being related to measurements in the same way.

 A pseudotensor $ t^{\mu}_{\nu}$ shares with the Einstein tensor $G^{\mu\nu}$ the physically interesting property of  having  a single recipe for inferring its components in a coordinate system from the metric components and their partial derivatives in that coordinate system. Thus the components do form a natural kind in that physical sense.   However, there is no transformation rule that allows one to infer the components with respect to one coordinate system from the components in another, so there is no mathematical unity. While this is generally taken to be a serious problem, it is in fact an essential virtue for representing infinitely many distinct energies.   The components of a pseudotensor with respect to different coordinate systems, being components of an object but not a geometric object, are just different entities, just as is required to describe the localization of different energy-momenta. While the possibility of writing down a pseudotensor in every coordinate system is occasionally discussed  \cite{EinsteinFoundation,PetrovChapter}, 
 the fact that the resulting collection is coordinate-invariant in a non-standard way and hence appropriate for representing the infinity of  gravitational energies seems never to have been noticed explicitly.

It is sometimes held  that modern differential geometry is or ought to be ``coordinate-free,''   and while one might need to ``introduce'' a coordinate system on certain  occasions, such occasions ought to evoke regret. The excellent text by Robert Wald \cite{Wald} is representative.  Thus  Wald describes any ``additional structure on spacetime, such as a preferred coordinate system or a decomposition of the spacetime metric into a `background part' and a `dynamical part','' which one would need to get a ``meaningful expression quadratic in first derivatives of the metric'' (as some famous pseudotensors are), as ``completely counter to the spirit of general relativity, which views the spacetime metric as fully describing all aspects of spacetime structure and the gravitational field.'' \cite[p. 286]{Wald}  In a moment of practical application, Wald manages to employ a pseudotensor anyway (pp. 84, 85),  though presumably without relish.  It is worth pointing out, however, that if there is any such thing as coordinate-free differential geometry, it isn't displayed in the bulk of Wald's book or most other literature where one might have thought to find it.  One ought to recall that \emph{all possible coordinate systems are already introduced in the definition of a manifold} \cite[p. 12]{Wald}.  Because the coordinate systems are already introduced in the greatest imaginable profusion at the start, there can be no objection to using them in the localization of gravitational energy-momentum.  The only possible objection (apart from possible difficulties in globalizing the results with bundle technology \cite{Misnerg00,ThirringForms}) can be to preferring some over others. 
The tacit assumption of the uniqueness of gravitational energy-momentum appears in the singular nouns:  ``a preferred coordinate system or a decomposition''  \cite[p. 286]{Wald}. 
 Obviously the infinite-component entity constructed from a pseudotensorial expression in all coordinate systems avoids preferring any particular coordinate system or class thereof over others.   Apparently there just isn't any ``coordinate-free'' way to express gravitational energy-momentum localization without  auxiliary objects besides the metric.  Coordinate systems are one option, and they are already present  anyway.  The natural conclusion is that certain aspirations to mathematical elegance and economy are chimerical, but relaxing excessively strict standards lets a solution appear.  %just not achievable.  
If one is committed to avoiding the use of any coordinate basis components (and hence avoiding using all of them) in favor of a formally ``coordinate (basis) free'' presentation, one can take all possible bases of commuting tangent vector fields as the relevant auxiliary structures.  A basis of commuting tangent vector fields is just the natural basis for a chart \cite[p. 27]{Wald} \cite[p. 471]{JMLee} by another name, so all  possible bases of commuting tangent vectors are just the natural bases for all possible charts.

%%%%%%%%%%%%%%%%%%%%%%%%%%%%%%%%%%%%%%%%%%%%%%%%%%%%%%%%%%%
\section{Energy-Momentum Localization as Infinite-Component Geometric Object in One Metric}

If one does not wish to express the metric $g_{\mu\nu}$ in terms of every coordinate system, but rather to express it in just one coordinate system, then that goal can be achieved in a certain sense.  The bimetric formalism gives an easy path to the result.  Whereas above the coordinates were fixed so that the flat metrics all took the form $diag(-1,1,1,1)$ while the components of the curved metric took various forms, one can instead fix the coordinates so that the curved metric takes a single form while the flat metrics take various forms.  These various forms will all look like $ \frac{ \partial x^M}{\partial y^{\alpha}} \eta_{MN} \frac{ \partial x^N}{\partial y^{\beta}}$ for all possible coordinate transformations $\frac{ \partial x^M }{\partial y^{\alpha}},$ with the  analogous result for a flat background connection.  The resulting infinite-component collection singles out some specific coordinate system as primary for expressing the metric $g_{\mu\nu},$ while also making reference to all other coordinate systems.  The result is the set of  components of an infinite-component geometric object in the chosen coordinate system.

%%%%%%%%%%%%%%%%%%%%%%%%%%%%%%%
\section{Objections to Pseudotensors Wrongly Assume Uniqueness of Energy}

Having developed the covariant construction of localized energy-momenta, one can now easily resolve some standard objections to pseudotensors, which already appeared in Pauli's review \cite{Pauli} and have reappeared in countless places since then.  For example, it is noted with disappointment that a given pseudotensor (at least one without second derivatives) can be made to vanish at any point or along any worldline by a suitable choice of coordinates.  With the tacit assumption that gravitational energy-momentum is unique, one then concludes that there is no real fact of the matter pertaining to the density of gravitational energy-momentum at that point or along that worldline.  But the point or worldline was arbitrary, so there is no fact of the matter about gravitational energy-momentum localization in general.  (Sometimes it is held that the situation  improves somewhat when symmetries yield Killing vectors, as in the case of spherical symmetry \cite[p. 603]{MTW}.)  It is now clear how this objection goes astray:  the components of a given pseudotensor with respect to different coordinate systems in fact pick out \emph{different energies}, some but not all of which vanish at the arbitrarily chosen point or along the arbitrarily chosen worldline.  The fact that some energies vanish there but others don't is a bit unfamiliar, but it is in no way paradoxical on reflection.

Given long disappointment with gravitational energy localization, many authors have turned to seeking quasilocalization, in which the energy in some volume is specified, rather than the energy density at a point. Quasilocal energy is generally expected to be unique.  The injustice of that expectation, however, follows from the multitude of local energy densities pointed out by Bergmann \cite{BergmannConservation}.   Pseudotensors are related to quasilocal methods \cite{NesterQuasiPseudo,NesterQuasi}.   It is sometimes expected that a good quasilocal mass (energy) should vanish in flat spacetime, though that criterion does not hold for every proposed definition \cite{Bergqvist}.  Likewise positive definiteness is  sometimes expected, though not always achieved \cite{Bergqvist,SzabadosReview}. Local  gravitational energy-momentum expressions do not reliably vanish in Minkowski space-time for all gauges either; instead they vanish in some coordinate systems (or some gauges \cite{PetrovChapter}) but not others.  If this result seems problematic, the resolution, again, is to notice that different coordinate systems/gauges pick out different energies.  It is a bit surprising that some of them fail to vanish even in Minkowski space-time, but it is not absurd.  Minkowski space-time is perhaps  unusual in that \emph{there exists} an energy-momentum density that vanishes everywhere. In Minkowski spacetime some energy densities will not vanish, but will integrate to vanishing total mass-energy; if the curved metric differs from the flat metric (or matrix $diag(-1,1,1,1)$ solely due to some localized gauge transformation, then such a situation should arise \cite{Grishchuk}.  If the total energy can vanish for an energy density that does not vanish everywhere, then positivity must also fail.  It appears, then, that both vanishing for Minkowski spacetime and positive definiteness are excessively strong conditions to impose on all of the infinitely many energies in a gauge-invariant localization, whether local or quasilocal.  The existence of one such energy (out of the infinite-component complex) with such properties is not ruled out, of course.   If one could find some way to restrict the auxiliary structures  to take notice of any Killing vectors (or commuting ones at least) of the metric $g_{\mu\nu}$ and adapt the coordinates accordingly, then a gauge-covariant energy-momentum expression that vanishes in flat spacetime might perhaps be devised; it would no longer be necessary to admit all coordinate systems in order to achieve gauge invariance. The proposal that energy localization makes sense in General Relativity just in case there is spherical symmetry \cite[p. 603]{MTW} is a variant, albeit too restrictive, of this idea.

Concerning Bauer's objection that flat spacetime in unimodular spherical coordinates has nonzero Einstein pseudotensor energy density \cite{BauerEnergy,Pauli}, the fact that the same  pseudotensorial expression in  different coordinate systems picks out different energies removes the paradox. The fact that the total energy in these spherical coordinates diverges \cite[p. 176]{Pauli} is not terribly  surprising, given that spherical coordinates have marvelously strong coordinate effects. 
Due to  the unimodular condition $\sqrt{-g}=1,$ the components of the metric tensor $g_{\mu\nu}$ tend to vanish or diverge at the origin and also at infinity; the inverse metric and their derivatives inherit comparable bad behavior.  (The unimodular condition is not too important apart from the details of this sort of misbehavior.)   Furthermore, spherical coordinates  are not well-defined everywhere that a corresponding set of Cartesian coordinates is defined, such as at the origin, so it is not  clearly meaningful (especially without introducing bundle techniques) to calculate the energy of all space in spherical coordinates.  The problem here seems to lie more with a poorly formulated  question than with an absurd answer.

Another traditional objection, this one due to Schr\"{o}dinger, calls attention to the vanishing of an Einstein pseudotensor (outside the Schwarzschild radius) for the Schwarzschild space-time in nearly Cartesian coordinates  with the unimodular condition $\sqrt{-g}=1$ \cite{SchrodingerEnergy,Pauli}.  Once again the existence of many distinct energy densities is helpful to recognize.  Possibly one would expect the \emph{total} mass-energy to come out ``right''  in this context, but various localizations are known to exist, in some cases with the energy all in some small region, in others not \cite{PetrovPoint,PetrovChapter}. If Schr\"{o}dinger had shown that \emph{all} the gravitational energies vanished outside the Schwarzschild radius, such a  result might be worrisome, but no such thing was shown.  That his particular energy vanishes is an interesting feature of gravitational energy as defined by the Einstein pseudotensor, but it is no real objection. In short,  traditional objections to pseudotensors are unpersuasive once faulty assumptions, especially the assumption of uniqueness, are cleared away.

In the last decade or two there has been in some circles a renewed interest in pseudotensors as yielding physically meaningful energy-momentum localizations  (\emph{e.g.}, \cite{RosenVirbhadra,Xulu}).  While the calculation of energy-momentum distributions using a finite collection of pseudotensors for a finite collection of metrics in a finite collection of coordinate systems might give  suggestive results, no answer to the gauge dependence problem could be achieved.  Gauge invariance requires using all admissible coordinate systems.  

%%%%%%%%%%%%%%%%%%%%%%%%%%%%%%%%%%%%%%%%%%%%%%%%%
\section{Comparison to Komar Energies and Schutz-Sorkin Noether Operator}

It is a somewhat familiar  point that there are infinitely many gravitational energies \cite{BergmannConservation,KomarConservation,Anderson,ReggeTeitelboim,FatibeneFrancaviglia}, though this fact has not had the influence that it ought to have, even on some of these authors.   As noted above, some of the energies might  vanish, while others might in some contexts be regarded as faces of the same entity.  Every vector field yields a Komar  `energy' flux (using the term broadly enough to ignore whether the vector is timelike), so one has a gravitational energy-momentum operator that is differential, not algebraic
as with simpler field theories, in its operation on the vector field.  If the Komar expressions were satisfactory, then a family of tensorial energies based on a family of vector fields would be a suitably covariant result depending in a fairly minimal way on auxiliary structures.  However, it is known that the Komar expression gives the wrong values for global conserved quantities in key cases, such as the ``factor of 2'' mismatch between mass and angular momentum for the Kerr solution \cite{KatzKomar,Katz,PetrovKatz2,FatibeneFrancaviglia,SzabadosReview}. %BakJackiw
 One of the most basic tasks of an energy-momentum localization is surely the derivation of correct global conserved quantities.  Thus the Komar expression cannot be  correct.  Given the uniqueness of the Komar result \cite{PavelleKomar,PavelleCurl}, the right answer must be non-covariant or, alternatively, depend on additional auxiliary structures.  An obvious  choice is to use the natural basis  from a coordinate system.  A coordinate system $x^{\mu}$ yields a natural cobasis of exact covectors $\mathbf{d}x^{\mu}$ and its reciprocal natural basis $\vec{ \frac{\partial}{\partial x^{\mu} }  }$ of commuting tangent vectors.  One can feed a natural basis into the (typically non-tensorial) Noether operator \cite{SchutzSorkin,SorkinStress,SzabadosSparling,ThirringForms} to get components of a pseudotensor.  Thus the components of the Noether operator relative to all natural bases yield the same sort of infinite-component covariant object as was obtained above.

%%%%%%%%%%%%%%%%%%%%%%%%%%%%%%%%%%%%
\section{Logical Equivalence of All Conservation Laws to Einstein's Equations}

In a typical field theory, one achieves energy-momentum conservation by noting that every field present in the equations of motion either has  Euler-Lagrange equations or has generalized Killing vector fields in the sense of vanishing Lie derivative \cite{TrautmanUspekhi}. (The generic notion of Lie differentiation for geometric objects is known, but not very widely \cite{SzybiakLie,Tashiro1,Tashiro2}.) In General Relativity as typically formulated (without a background metric or connection), every field present has Euler-Lagrange equations; there are no non-variational fields (to borrow a useful term \cite{GotayIsenberg}).  One might then expect that the energy-momentum of matter and gravity together to be conserved using both the gravitational field equations and the matter field equations.  A distinctive feature of General Relativity is that, because of gravitational gauge invariance (see, \emph{e.g.}, \cite{SliBimGRG}), conservation follows using the gravitational field equations alone, without using the matter equations \cite{Anderson,Wald}. 
One can take the gravitational stress-energy  $t^{\mu}_{\nu} \sqrt{-g}$ to be  
\begin{equation} t^{\mu}_{\nu} \sqrt{-g} =_{def} - G^{\mu}_{\nu} \sqrt{-g}  - \mathcal{F}^{[\mu\alpha]}_{\nu},_{\alpha}, \end{equation}
where $\mathcal{F}^{[\mu\alpha]}_{\nu},_{\alpha}$ is an arbitrary expression apart from having identically vanishing divergence and being built from the metric components and their derivatives and perhaps some constant matrices such as $\eta_{MN}.$ (Newton's constant has been suppressed for convenience.)
Combining the gravitational and material stress-energies gives the total energy-momentum complex $ \frak{T}^{\mu}_{\nu} \sqrt{-g} =_{def} T^{\mu}_{\nu} \sqrt{-g}  +  t^{\mu}_{\nu} \sqrt{-g}.$ 
This total stress-energy complex satisfies a conservation law with a coordinate divergence:
\begin{eqnarray}  \frac{ \partial}{\partial x^{\mu}} ( \frak{T}^{\mu}_{\nu} \sqrt{-g}) = \frac{\partial}{\partial x^{\mu} } ( T^{\mu}_{\nu} \sqrt{-g}  +  t^{\mu}_{\nu} \sqrt{-g})= 
\frac{ \partial}{ \partial x^{\mu} } ( T^{\mu}_{\nu} \sqrt{-g}   - G^{\mu}_{\nu} \sqrt{-g}  - \mathcal{F}^{[\mu\alpha]}_{\nu},_{\alpha} ) \nonumber \\
= \frac{\partial}{ \partial x^{\mu}} ( T^{\mu}_{\nu} \sqrt{-g}   - G^{\mu}_{\nu} \sqrt{-g})  - \mathcal{F}^{[\mu\alpha]}_{\nu},_{\alpha\mu} 
= \frac{\partial}{\partial x^{\mu}} ( T^{\mu}_{\nu} \sqrt{-g}   - G^{\mu}_{\nu} \sqrt{-g}) 
=0, \end{eqnarray}
where Einstein's field equations have been used in the last line.  
One now sees that the total energy-momentum density vanishes (when Einstein's equations hold) except for a curl, for which reason one can calculate conserved quantities with a surface integral.  Einstein's equations entail a pseudotensorial conservation law in every coordinate system.  As has been observed above, the collection of pseudotensorial laws of a given form in all coordinate systems is invariant in an appropriate, although unfamiliar, sense.  

The collection of all of the pseudotensorial  conservation laws is in fact \emph{equivalent} to Einstein's equations \cite{Anderson}, so the reverse entailment also holds, as will now appear.  In any coordinate system, from the conservation law $\partial/\partial x^{\mu} ( T^{\mu}_{\nu} \sqrt{-g}  +  t^{\mu}_{\nu} \sqrt{-g})=0$ (with $t^{\mu}_{\nu} \sqrt{-g}$ defined as above), one obtains 
  \begin{equation} T^{\mu}_{\nu} \sqrt{-g} - G^{\mu}_{\nu} \sqrt{-g} =  \mathcal{F}^{[\mu\alpha]}_{\nu},_{\alpha} \end{equation} for some $\mathcal{F}^{[\mu\alpha]}_{\nu},_{\alpha}$ (which might, for all that has appeared so far, vary from one coordinate system to another, because of the arbitrary curl that one can include in $\mathcal{F}^{[\mu\alpha]}_{\nu},_{\alpha},$ like a constant of integration).  But the left side is a tensor density, so the right side must be one also.  There is no nonzero tensor density that is built out of the allowed ingredients and that has the right number of indices, so the right side must be zero.  Thus the totality of the pseudotensorial conservation laws indeed entails Einstein's equations 
$T^{\mu}_{\nu} \sqrt{-g} - G^{\mu}_{\nu} \sqrt{-g} = 0.$   The fact that the conservation laws entail the field equations sheds light on those approaches to General Relativity that aim  to derive the field equations using the conservation laws as premises or lemmas \cite{EinsteinEntwurf,Deser,SliBimGRG}.

Pseudotensor conservation laws routinely have been accused of being physically meaningless on account of vicious dependence on a choice of  coordinates.  While some authors emphasize the physical reality  of gravitational radiation that can heat or move objects, notwithstanding mathematical difficulties \cite[pp. xxvi, 219]{Feynman} \cite{HoleStory}, or choose to accentuate the positive features of pseudotensors \cite{Weinberg}, others emphasize mathematical purity and belittle the conservation laws \cite{MTW,Hoefer}.   It is now clear that there is no  vicious coordinate dependence.  Even apart from recognizing that fact,  it is highly doubtful that anything physically meaningless is logically equivalent to Einstein's equations.  This logical equivalence is another way of recognizing that the set of all pseudotensor conservation laws is indeed gauge invariant and hence physically meaningful.  This section could be repeated with insignificant changes to show the logical equivalence of the collection of all bimetric conservation laws (each of which is gauge dependent) to Einstein's equations, thus showing the physical significance and gauge invariance of the whole collection. One might stop short of saying that General Relativity \emph{just is} local energy-momentum conservation, but it is difficult to imagine a greater disagreement than that between the usual claim that General Relativity does not support a law of local energy-momentum conservation and the  mathematical fact that Einstein's equations are logically equivalent to a gauge-invariant infinite-component local energy-momentum conservation law.  
 From this point of view it is obvious that there is a connection between the first law of thermodynamics and Einstein's equations.

%%%%%%%%%%%%%%%%%%%%%%%%%%%
\section{Angular Momentum Localization}

A suitably covariant localization of gravitational energy-momentum was obtained above by collecting together the pseudotensors of a given type (such as that of Einstein or of Landau and Lifshitz or one of Goldberg's \cite{GoldbergConservation}) in every coordinate system.  The resulting pseudotensors can depend on the auxiliary matrix $diag(-1,1,1,1),$ though some choices do not. Those that do not \cite{GoldbergConservation} tend to behave worse regarding angular momentum than those that do  \cite{ChangNesterChen}.  
 For angular momentum, one  introduces the coordinates $x^{\mu}$ and a symmetric choice of total energy-momentum complex $ \sqrt{-g} \frak{T}^{\mu\nu}  $  so that
\begin{equation} \frak{M}^{\mu\nu\alpha}=_{def}  \sqrt{-g} \frak{T}^{\mu\nu}  x^{\alpha} - \sqrt{-g} \frak{T}^{\mu\alpha}  x^{\nu}
\end{equation}
satisfies the conservation law
\begin{equation} \frac{ \partial}{\partial x^{\mu}} \frak{M}^{\mu\nu\alpha}=0 \end{equation} 
because of $ \frac{ \partial}{\partial x^{\mu}} (\sqrt{-g} \frak{T}^{\mu\nu})=0.$ By parity of reasoning with the above, the collection of these angular momentum densities in \emph{every} coordinate system is an appropriate covariant infinite-component object.  Thus angular-momentum achieves a gauge-invariant localization in the same way as energy-momentum.  
If flat background metrics are used instead of coordinate systems, then position $4$-vectors take the place of the coordinates.  Of course any  non-uniqueness of the gravitational energy-momentum (pseudo-)tensor due to relocalization by adding a curl will have consequences  for angular momentum localization.

%%%%%%%%%%%%%%%%%%%%%%%%%%%%%%%%%%%%%%%

\section{Conceptual Benefits of Energy  Localization and Conservation}

One reason for seeking a conservation law for gravitational energy-momentum is to characterize the properties of various gravitational fields or space-times.  Komar wrote some time ago of part of his own work that
\begin{quote}
all of the above  attempts to generalize energy and momentum to arbitrarily curved manifolds are formal in character, and the ``correct'' choice, if indeed there is one, must be determined by the use to which we wish to put the resulting conservation laws.  One is not particular interested in a formal definition of energy if it teaches us nothing about the properties of the spaces under consideration. \cite[p. 1413]{KomarAsymptotic}
\end{quote}
If one is  aware of the uses to which the supposed lack of an energy conservation law in General Relativity has been put by now, however, then the benefits of even a formal local energy conservation law become evident.  
The received view that there is no gauge-invariant and hence physically meaningful local conservation law for energy-momentum in General Relativity tends to inspire (though not strictly entail) a variety of unwarranted  conclusions. 
Some have criticized or rejected General Relativity (or Big Bang cosmology in particular) as having mystical tendencies on account of its supposed lack of conservation laws, while others have  appealed to General Relativity for certain purposes for the same reason.  
It has been claimed, to be specific, %, in every case by a physicist or philosopher of physics,
 that the lack of a local conservation implies: % list names if cited by number rather than author-date
%\vspace{-.1in}
\begin{enumerate} 
  \item that General Relativity is false  (by A. A. Logunov and collaborators \cite{LogunovEnergy1977,LogunovExplain}, addressed in \cite{FaddeevEnergy,Zel1,Zel2,Grishchuk90}); 
  \item  that Big Bang cosmology violates energy conservation and so is false (by Robert Gentry, addressed in \cite{Gentry2,Gentry3}); 
  \item that Big Bang cosmology is plausibly true and yet violates energy conservation, which  is so fundamental as to transcend physics into metaphysics (by Mario Bunge \cite{BungeEnergy});  the tension seems not to be noticed;
  \item that Big Bang cosmology  violates energy conservation and so is a useful heat sink for anomalous terrestrial heat production (by D. Russell Humphreys, addressed in \cite{HumphreysConserve}); 
  \item that General Relativity makes it easier than do other field theories for immaterial souls to affect bodies (by Robin Collins \cite{CollinsEnergy}); % see also \cite{EnergyMental}); 
  \item and that universes with zero total energy can come into being without violating energy conservation (by Edward Tryon \cite{Tryon} and Walter Thirring \cite{ThirringCreation}).
  \end{enumerate}     Concerning the last claim,  once the gauge-invariant local conservation of  energies is recognized, it is clear that only universes for which  all the uncountably many energies vanish could pop into existence without violating energy conservation, a condition that is difficult or impossible to satisfy.  While these six  conclusions are seen not to follow  when the effort to produce a sufficiently detailed and subtle analysis is made, %and defenders of standard views continue to control the bulk of relevant academic publishing,  
unwarranted conclusions continue to arise because the knock-out blow to forestall them, namely, a satisfactory local conservation law including gravitational energy, is incompatible with the received view of that subject. 
Larry Laudan has argued that scientific progress can occur not only by solving empirical problems, but also by solving conceptual problems \cite{LaudanProgress}.   
Identifying gauge-invariant and hence physically meaningful local conservation laws  therefore contributes to scientific rationality by resolving a conceptual problem in General Relativity.

%%%%%%%%%%%%%%%%%%%%%%%%%%%%%%%%%%%%%%

\section{Conclusion}
It is ironic that though Einstein used energy conservation as a  criterion for finding his field equations, it was widely concluded that  the resulting theory lacked any local conservation law for energy-momentum. That irony is resolved by recognition of an \emph{infinite-component gauge-invariant} local energy-momentum conservation law, which is shown to follow from Einstein's equations.  Local energy-momentum conservation, as Anderson noted, is even logically equivalent to Einstein's equations. The principle that the real is the invariant, characteristic of 20th century mathematical physics \cite{HowardIdeal}, has appeared to be an obstacle to the reality of gravitational energy localizations until 
now. The principle does not say with respect to what invariance is desired or how it should manifest itself, but generally accepted background assumptions provided an answer, albeit a flawed one.   Pseudotensors are not invariant in the sense traditionally expected, so the corresponding local conservation laws have been widely viewed as mere mathematical artifices.  Now that the appropriate sense of invariance has been noticed and the connection to the existence of infinitely many gravitational energies has been recognized, there is no difficulty in regarding the whole infinite family of gravitational energy localizations (as picked out by some specific pseudotensorial functional form or the like) as gauge-invariant and hence real.  Thus the conceptual problem of gauge-dependence of gravitational energy-momentum is solved.

The question of gauge-dependence of gravitational energy-momentum localization is largely orthogonal to the question of getting the `right answers' for the conserved quantities (except for disqualifying the Komar expression), a matter of the technical details of the specific choice of pseudotensor.  For that project, one needs to choose an appropriate functional dependence on the metric and auxiliary structures.   One would expect gravitational energy-momentum to have (besides the now-resolved gravitational gauge dependence problem) all the usual non-uniqueness (even on-shell in some cases) of  symmetrizing,  improving \cite{ImprovedEnergy}, relocalizing by curls in general \cite{Anderson,LeclercEnergy}, making  field redefinitions \cite{PonsEnergy}, and the like, presented by other field theories. There might be ways to tame that non-uniqueness either for every solution in the same fashion \cite{KatzBicakLB} or on a case-by-case basis   \cite{NesterQuasiPseudo,NesterQuasi}.   A  satisfactory treatment of gravitational energy ought to be achievable by taking the  best functional form  on technical grounds and rendering it covariant it in the way outlined here.

\section{Acknowledgments}

I thank Don Howard and Alexander N. Petrov for conversations and  Katherine Brading for discussion of  the role of energy conservation in Einstein's process of discovery of General Relativity.

%\bibliography{Pitts}  

%\bibliographystyle{unsrt}  %apalike} 
%%
%%
\end{document}